\newif\ifcarlo
\carlofalse

\documentstyle[aps,twocolumn,epsfig,floats]{revtex}

\begin{document}

\draft

\date{November 1999}
\title{Search for two-scale localization in disordered
wires in a magnetic field}
\author{H. Schomerus and C. W. J. Beenakker}
\address{Instituut-Lorentz, Universiteit Leiden, P.O. Box 9506, 2300 RA
Leiden, The Netherlands}

\ifcarlo
\relax
\else
	\twocolumn[
\fi
\widetext
\begin{@twocolumnfalse}

\maketitle

\begin{abstract}

The supersymmetry technique
\lbrack{}A. V. Kolesnikov and K. B. Efetov,
Phys.\ Rev.\ Lett.\ {\bf 83}, 3689 (1999)\rbrack{} predicts
a two-scale behavior of wavefunction decay
in disordered wires in the crossover regime  from preserved to
broken time-reversal symmetry.
We have tested this prediction by 
a transmission approach, relying on the
Borland conjecture that relates the decay length of the transmittance
to the decay length of the wavefunctions.
Our numerical simulations show no indication of two-scale behavior.

\end{abstract}


\vspace{0.5cm}

\narrowtext

\ifcarlo
\relax
\else
\end{@twocolumnfalse}
]
\fi
\narrowtext

In a remarkable paper \cite{KE},
Kolesnikov and Efetov have predicted that the decay of
wavefunctions in disordered wires is characterized by two localization
lengths, if time-reversal symmetry is partially broken by a weak magnetic
field. Using the supersymmetry technique \cite{efetov} it was demonstrated that the
far tail of the wavefunctions decays with the length $\xi_2$
characteristic for completely broken time-reversal symmetry---even if the flux
through a localized area is much smaller than a flux quantum. At
shorter distances the decay length is $\xi_1=\frac{1}{2}\xi_2$. It was
suspected that previous studies by
Pichard et al.\ \cite{Pichard} 
found single-scale decay
because of the misguiding theoretical
expectation of such behavior.
This expectation was also the basis for the interpretation of the 
experiments by Khavin, Gershenson, and Bogdanov \cite{Khavin}
on submicron-wide wires.

The prediction of Kolesnikov and Efetov calls for a test by means of a
dedicated experiment  or computer simulation. It is the purpose of this
work to provide the latter. We target the key feature of the two-scale 
localization phenomenon, which is the doubling of the asymptotic decay
length at infinitesimally weak magnetic fields.

Our numerical simulations are based on a transmission approach.
We rely on the Borland conjecture \cite{Borland} (believed 
to be true generally \cite{prm}), that relates the
asymptotic decay of the
transmittance $T$ with increasing wire length $L$
to the asymptotic decay of the wavefunction $\psi(L)$. According to the 
Borland conjecture, 
the Lyapunov exponent $\alpha=-\lim_{L\to \infty} \frac{1}{2} L^{-1}
\ln T$ is identical to the
inverse localization length
$\xi^{-1}=-\lim_{L\to \infty}L^{-1}\ln |\psi(L)|$.
Moreover, $\xi$ and $\alpha$ are self-averaging, meaning that the
statistical fluctuations become smaller and smaller as $L\to\infty$.
Our numerical simulations show that the crossover from $\xi=\xi_1$ to
$\xi=\xi_2$ does not occur until the flux $\Phi_\xi$ through a wire
segment of length $\xi_1$ is of the order of a flux quantum $\Phi_0=h/e$.
For our longest wires ($L\gtrsim 150\xi_1$) the crossover according to
Ref.\ \cite{KE} should have occurred at $\Phi_\xi/\Phi_0\simeq
\exp(-L/8\xi_1)\simeq 10^{-8}$.
We consider various possible
reasons for the disagreement with Ref.\ \cite{KE} (finite number of
modes, anomalously localized states), but believe that none of these
provides a satisfactory explanation.

\begin{figure}
\epsfig{file=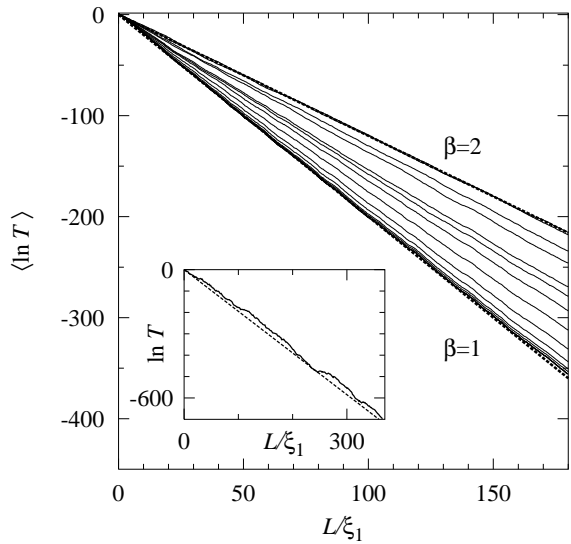,width=3.2in}
\medskip
\caption{Average logarithmic transmittance $\langle \ln T\rangle$
as a function of
wire length $L$ for the Anderson model with
$N=5$ propagating modes.
The two dashed lines have the slopes predicted for
preserved ($\beta=1$) and broken ($\beta=2$) time-reversal symmetry.
From bottom to top the data
corresponds to fluxes $\Phi_\xi/\Phi_0$=0, 0.0005, 0.005,
0.05 (four indistinguishable solid curves),
0.5, 1, 2.5, 5, 10, 15, 20, 25, 40,
50, 75, 125 (two indistinguishable solid curves).
The inset shows $\ln T$ for an individual
realization with $\Phi_\xi=\frac{1}{2}\Phi_0$ (solid curve)
and the slope of the ensemble-averaged result (dashed
line).
}
\label{fig:1}
\end{figure}

Our first set of results is obtained from the numerical calculation
(by the technique of recursive Green functions
\cite{recgf4})
of the
transmission matrix $t$ for a two-dimensional Anderson Hamiltonian with
on-site disorder.
In units of the lattice constant $a\equiv 1$, the width of
the wire is $W=13$ and the wavelength of the electrons is 
$\lambda=5.1$, resulting in $N=5$ propagating modes through the wire.
The localization lengths $\xi_1=(N+1) l$ and $\xi_2=2 N l$
are determined by the scaling parameter $l$
of quasi-one dimensional localization theory, which differs from the transport
mean-free path by a coefficient of order unity \cite{BeenRMP}.
The average of the transmittance $T=\mbox{tr}\, t t^\dagger$
in the metallic regime, fitted to
$\langle T\rangle=N(1+L/l)^{-1}$,  yields $l=65$.
This gives
a localization length $\xi_1= 390$
for preserved time-reversal symmetry (symmetry index $\beta=1$)
and a localization length $\xi_2=650$
for broken time-reversal symmetry ($\beta=2$).

Fig.\ \ref{fig:1} shows the ensemble-averaged logarithm of the
transmittance
$\langle \ln T\rangle$ as a function of wire
length $L$ for various values of the magnetic field $B$
(or flux $\Phi_\xi=W \xi_1 B$).
We find a smooth
transition between the theoretical expectations for
preserved and broken time-reversal symmetry.
Most importantly, we find an asymptotic slope
$s(B)=\lim_{L\to\infty}L^{-1}\langle \ln T \rangle$
that interpolates smoothly
between the values $s=-2/\xi_1$ for $B=0$ and $s=-2/\xi_2$ for large
$B$.
There is no indication of a crossover
to the slope $s=-2/\xi_2$ for smaller values of $B$,
even for very long wires ($L\gtrsim 150\xi_1$).
According to the theory of Ref.\ \cite{KE}, the crossover should occur
at a length $L_{\rm cross}$ given by
\begin{equation}
\label{eq1}
L_{\rm cross}/\xi_1=8\ln (\sqrt{12}\Phi_0/4\pi\Phi_\xi)+{\cal O}(1),
\end{equation}
which is well within the range of
our simulations ($L_{\rm cross}\simeq 14\xi_1$ for
$\Phi_\xi\simeq 0.05 \Phi_0$).
The absence of two-scale behavior in the transmittance of
an individual, arbitrarily chosen realization
is demonstrated in the inset of Fig.\ \ref{fig:1}, for $\Phi_\xi=\case 12 \Phi_0$.
The self-averaging property of the Lyapunov exponent is evident.

\begin{figure}
\epsfig{file=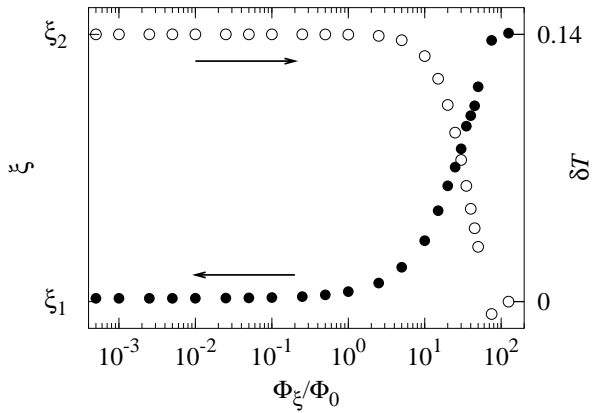,width=3.2in}
\medskip
\caption{Asymptotic decay length 
(solid circles) and weak-localization correction 
$\delta T$ (open circles)
as a function of flux for the $N=5$ Anderson model.
}
\label{fig:2}
\end{figure}

The asymptotic decay length $\xi(B)=-2/s(B)$ is plotted
versus magnetic field in Fig.\ \ref{fig:2}, together with the
weak-localization correction $\delta T=T(B=\infty)-T(B)$ at $L=\xi_1$. 
For both quantities, breaking of
time-reversal symmetry sets in when
$\Phi_\xi$ 
is comparable to $\Phi_0$.
The transition from $\beta=1$ to $\beta=2$ is completed
for $\Phi_\xi\approx 100 \Phi_0$.

Our second set of results is obtained from a 
computationally more efficient model
of a disordered wire, consisting of a chain of chaotic cavities
(or quantum dots) with two leads attached on each side.
This so-called `domino' model \cite{domino} is similar
to Efetov's model of a granulated metal \cite{efetov} and to the
Iida-Weidenm\"uller-Zuk model of connected slices \cite{IWZ}.
The length $L$ is now measured in units of cavities, and the mean free
path $l=1$.
The scattering matrices
of each cavity are randomly drawn from an ensemble
(proposed by \.Zyczkowski and Ku\'s \cite{ZK})
that interpolates (by means of a parameter $\delta$)
between the circular orthogonal ($\beta=1$, $\delta=0$)
and unitary ($\beta=2$, $\delta=1$) ensembles of random-matrix theory.
The relationship between $\delta$ and $\Phi_\xi/\Phi_0$ is linear for
$\delta\ll 1$.

\begin{figure}
\epsfig{file=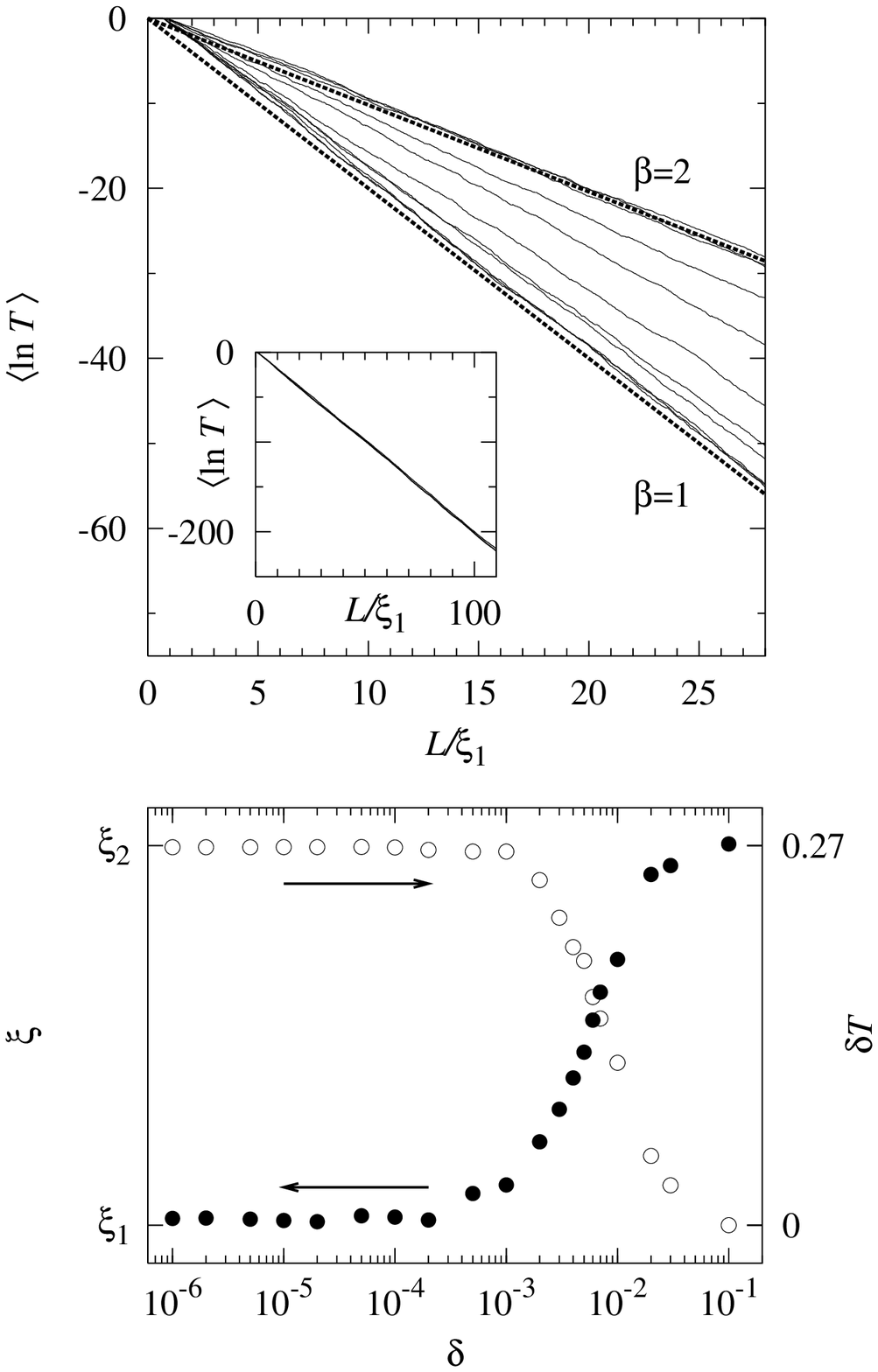,width=3.2in}
\medskip
\caption{Same quantities as in Figs.\ \protect\ref{fig:1} and
\protect\ref{fig:2}, but now for the $N=50$ domino model.
In the upper panel, the magnetic flux parameter
$\delta=0$, 0.0001, 0.0002, 0.0005, 0.001, 0.002, 0.005,
0.01, 0.02, 0.05, and 0.1. In the inset, $\delta=0$, 0.00001, and 0.0001
(indistinguishable curves).
}
\label{fig:3}
\end{figure}

We increased the number of propagating modes to $N=50$, because it is
conceivable
that the two-scale localization
becomes manifest only in the large $N$-limit, or
that only in this limit
the critical flux $\Phi_\xi$ for the transition from $\xi_1$ to $\xi_2$
becomes $\ll \Phi_0$.
(In the experiments
of Ref.\ \cite{Khavin} $N\approx 10$,
so our simulations are in the experimentally relevant range of $N$.)
Because of the much larger value of $N$, we restricted
ourselves for larger values of the 
magnetic flux to $L\simeq 25\xi_1$, which should be sufficient
to observe the localization length $\xi_2$ for $\Phi_\xi/\Phi_0\gtrsim
10^{-2}$. For smaller values of the flux, we increased the wire length to
$L\simeq 100 \xi_1$.
The data is presented in Fig.\ \ref{fig:3}.
It is qualitatively similar to the results for the $N=5$ Anderson model.
Instead of two-scale behavior, we only see a single decay
length which crosses over smoothly from $\xi_1$ to $\xi_2$ with
increasing $\delta$. Again, the crossover of $\xi$ coincides with the
crossover of the weak-localization correction, so there is no
anomalously small crossover flux for the localization length.

\begin{figure}
\epsfig{file=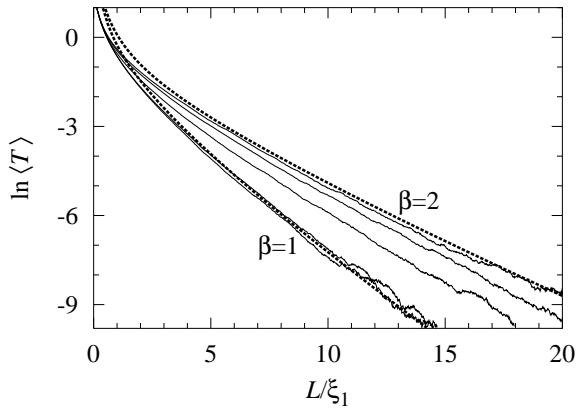,width=3.2in}
\medskip
\caption{Logarithm of the average transmittance
$\ln \langle T\rangle$
as a function of
wire length $L$ for the $N=5$ Anderson model 
at various values of the magnetic field (solid curves; from bottom to
top, $\Phi_\xi/\Phi_0=0,5,25,50,125$).
The dashed curves are the theoretical prediction of 
Refs.\ \protect\cite{Zirnbauer,MGZ} for zero and large magnetic field.
}
\label{fig:4}
\end{figure}

The logarithmic average $\langle \ln T\rangle$ is the experimentally 
relevant quantity since it is representative for a single 
realization (see Fig.\ \ref{fig:1}, inset). The average transmittance $\langle
T\rangle$ itself is not representative, because it is dominated by rare
occurrences of anomalously localized states \cite{Mirlin}.
Since Kolesnikov and Efetov \cite{KE} studied the average of
wavefunctions themselves, rather than the average of logarithms of
wavefunctions, it is conceivable that their findings are the result of
such rare occurrences. For completely broken or fully preserved
time-reversal symmetry the average transmittance is given by
\cite{Zirnbauer}
\begin{equation}
\label{eq:g}
\ln \langle T\rangle =-L/2\xi_\beta - \case{3}{2}\ln L/\xi_\beta
+{\cal O}(1).
\end{equation}
The order 1 terms are also known \cite{Zirnbauer,MGZ} and contribute
significantly for $L\lesssim 30\xi_1$.
(This is the numerically accessible range,
because anomalously localized states 
become exponentially rare with increasing wire length.)
We have plotted the full
expressions in Fig.\ \ref{fig:4} (dashed curves), together with the
numerical data for the $N=5$ Anderson model.
Again we find a smooth crossover
between preserved and broken time-reversal symmetry.  
There is no transition with increasing wire length to
a behavior indicative of completely
broken time-reversal symmetry,
even though the flux $\Phi_\xi$ is much larger
than required [according to Eq.\ (\ref{eq1})]
to observe this crossover for the wavefunctions.

In conclusion, we have presented a numerical search for the two-scale
localization phenomenon predicted by Kolesnikov and Efetov \cite{KE},
with negative result: The asymptotic decay length of the transmittance is
found to be given by $\xi_{1}$ and not by $\xi_{2}$, as long as the flux
through a localization area is small compared to the flux quantum. How can one
reconcile this numerical finding with the result of the supersymmetry
theory? We give three possibilities. (1) One might abandon the Borland
conjecture and permit the asymptotic decay length of the transmittance
(Lyapunov exponent) to differ from the asymptotic decay length of the
wavefunction (localization length). Since the Borland conjecture has been
the cornerstone of localization theory for more than three decades, this
seems a too drastic solution. (2) One could argue that the wires in the
simulation are too narrow or too short --- although they are in the
experimentally relevant range of $N$ and $L$. (3) One could attribute
the two-scale localization phenomenon to anomalously localized states, that
are irrelevant for a typical wire. We can not fully exclude these two
remaining possibilities, but they would severely diminish the experimental
relevance of the phenomenon.

A discussion with P. G. Silvestrov motivated us to look into this
problem. We acknowledge helpful correspondence with A. V. Kolesnikov and
support by the Dutch Science Foundation NWO/FOM.

\ifcarlo
\end{@twocolumnfalse}
\fi

\end{document}